\documentclass[dvips]{article}
\usepackage{graphicx}
\usepackage{amssymb}
\usepackage{graphicx}
\usepackage{dcolumn}
\usepackage{amsmath}
\usepackage{lscape}
\usepackage{longtable}

\begin{document}
\textwidth 16cm
\newcommand{\bd}{\begin{document}}
\newcommand{\ed}{\end{document}}
\newcommand{\bc}{\begin{center}}
\newcommand{\ec}{\end{center}}
\newcommand{\bfr}{\begin{flushright}}
\newcommand{\efr}{\end{flushright}}
\newcommand{\lt}{\left}
\newcommand{\rt}{\right}
\newcommand{\vs}{\vspace}
\newcommand{\hs}{\hspace}
\newcommand{\beq}{\begin{equation}}
\newcommand{\eeq}{\end{equation}}
\newcommand{\lb}{\linebreak}
\newcommand{\pb}{\pagebreak}
\newcommand{\mb}{\makebox}
\newcommand{\fb}{\framebox}
\newcommand{\mc}{\multicolumn}
\newcommand{\ben}{\begin{enumerate}}
\newcommand{\een}{\end{enumerate}}
\newcommand{\bit}{\begin{itemize}}
\newcommand{\eit}{\end{itemize}}
\newcommand{\ol}{\overline}
\newcommand{\un}{\underline}
\newcommand{\lefq}{\lefteqn}
\newcommand{\ba}{\begin{array}}
\newcommand{\ea}{\end{array}}
\newcommand{\beqa}{\begin{eqnarray}}
\newcommand{\eeqa}{\end{eqnarray}}
\newcommand{\beqas}{\begin{eqnarray*}}
\newcommand{\eeqas}{\end{eqnarray*}}
\newcommand{\bfg}{\begin{figure}}
\newcommand{\efg}{\end{figure}}
\newcommand{\bds}{\begin{displaymath}}
\newcommand{\eds}{\end{displaymath}}
\newcommand{\btb}{\begin{tabbing}}
\newcommand{\etb}{\end{tabbing}}
\bc {\huge $\mathcal{PT}$ Symmetric Hamiltonian Model and Dirac
Equation in 1+1 dimensions } \ec

\vs{1cm}

\bc
{\it \"Ozlem Ye\c{s}ilta\c{s}{\footnote {e-mail : yesiltas@gazi.edu.tr}\\
Department of Physics, Faculty of Science, Gazi University,
06500 Ankara, Turkey\\
\vspace{.16cm}

}} \ec \vs{1cm}
\begin{abstract}
 \noindent In this article, we have introduced a $\mathcal{PT}$ symmetric non-Hermitian Hamiltonian model  which is given as  $\hat{\mathcal{H}}=\omega (\hat{b}^{\dag}\hat{b}+\frac{1}{2})+  \alpha (\hat{b}^{2}-(\hat{b}^{\dag})^{2})$ where $\omega$ and $\alpha$ are real constants, $\hat{b}$ and $\hat{b^{\dag}}$ are first order differential operators. The Hermitian form of the Hamiltonian $\mathcal{\hat{H}}$ is obtained by suitable mappings and it is interrelated to the time independent one dimensional Dirac equation in the presence of position dependent mass. Then, Dirac equation is reduced to a Schr\"{o}dinger-like equation and two new complex non-$\mathcal{PT}$ symmetric vector potentials are generated.    We have obtained real spectrum for these new  complex  vector potentials using shape invariance method. We have searched the real energy values using numerical methods for the specific values of the parameters.

\end{abstract} {keyword:}  PT symmetry, pseudo-Hermiticity, Dirac equation.

\section{Introduction}
The nature of quantization arises due to the symmetry of the
equations governing the physics. So, symmetry has long been known
as a powerful and computational topic in quantum mechanics.
Recently, there have been many studies of $\mathcal{PT}$ symmetric
non-Hermitian systems with real energy since the original work of
Bender and Boettcher \cite{Bender} and the literature on such
systems has expanded rapidly \cite{benders, Cannata, Znojil,
Levai, Bagchi, Quesne}. One of the key points of the investigation
was the $\mathcal{PT} $ symmetry generated by the product of the
parity, $\mathcal{P}$, and time,$ \mathcal{T}$ , linear and
anti-linear inversion operators $\mathcal{P}x\mathcal{P}=-x$,
$\mathcal{T}x\mathcal{T}=x$, $\mathcal{T}i\mathcal{T}=-i$. The
operator $\mathcal{T}$ is anti-linear because it changes the sign
of $i$. If $\mathcal{PT}$  symmetry of the Hamiltonian is
unbroken; eigenfunction of the operator $\mathcal{PT}$ is
simultaneously an eigenstate of Hamiltonian $H$, i.e.  $[H,
\mathcal{PT}]=0$. Later, it has been realized that the existence
of real eigenvalues can be associated with a non-Hermitian
Hamiltonian provided it is $\eta$-pseudo-Hermitian \cite{Mos}:
$\eta H=H^{\dag}\eta$ where $\eta$ is a Hermitian linear
automorphism which can be given as $\eta=(OO^{\dag})^{-1}$, $O$ is
a linear invertible operator. Here, the Hilbert space equipped
with the inner product $<.,\eta .>$ is identified as the physical
Hilbert space. And the observable $\Theta$ which is the element of
physical Hilbert space is related to the Hermitian operator
$\theta$ by means of a similarity transformation $\Theta=\rho^{-1}
\theta \rho$ where $\rho=\eta^{2}$. At the same time Bagchi and
Quesne have established that the twin concepts of
pseudo-Hermiticity and weak-pseudo-Hermiticity \cite{quesne1}.
Thus, the concept of pseudo-Hermiticity has attracted much
interest on behalf of physicists \cite{ahmed, zno, swanson, jin}.

In \cite{Vakarchuk1}, the Kepler problem solutions are
investigated in Dirac theory for the particle whose mass is
position dependent and the effective mass is given in the form of
a multipole expansion, existence of the bound states are discussed
in detail. Earlier, using a standard expansion of radial functions
as a different approach, Dirac equation spectrum was obtained for
the mixed potentials \cite{soff}. Results of \cite{Vakarchuk1}
which are about large quantum numbers not leading to inverse mass
and momentum independent energy is also consistent with those
found in \cite{quesne}. In \cite{Vakarchuk2}, Dirac matrices
$\hat{\alpha}, \hat{\beta}$ have space factors as $f, f_{1}$
functions where $f$ responsible of the deformation of the
Heisenberg algebra for the coordinates and momentum operators,
$f_{1}$ is responsible of a dependence of the particle mass on its
position. Exact solutions are found for the fermion in Coulomb
field with the function $f$ which depends on $r$ linearly while
the function $f_{1}$ depends on $r$ inversely. It is pointed out
that the spectrum results of \cite{Vakarchuk2} can be useful for
the nanoheterosystems. Similar arguments about the Dirac
oscillator with deformed commutation relations leading to the
existence of the minimal length of space can be found in \cite{Q}.

Latterly, non-Hermitian potentials for the fermions have been
studied in the literature \cite{Jia1, Jia2, royc, om, castro} and
fermion models interacting with $\mathcal{PT}$ symmetric
potentials in presence of effective mass have been attracted
interest \cite{sinha, mustafa, mazhar, sever2, santos, egri,
long}. In \cite{sever1}, the one-dimensional effective mass Dirac
equation bound states are studied within the interactions of
non-$\mathcal{PT}$-symmetric, and non-Hermitian, exponential type
potentials. Moreover, $(1+1)$ Dirac equation with
position-dependent mass (PDM) and complexified Lorentz scalar
interactions, is discussed through the supersymmetric quantum
mechanics \cite{omar}. More references about the complex
potentials and Dirac theory can be found in \cite{Refs}.

Also, pseudo-Hermitian interaction in relativistic quantum
mechanics is studied with the positive definite metric operator
$\eta$ calculations for the state vectors \cite{gupta}. Using the
spin and pseudo-spin concept, spectrum of $PT$ symmetric
Rosen-Morse potential is studied and analytical methods are used
in \cite{sam}. Dirac equation with position dependent effective
mass transformed into Schr\"{o}dinger-like equation is studied in
a general context and L\'{e}vai's method is used \cite {pan}.
Supersymmetric quantum mechanics (SUSY QM) provides elegant
procedures to solve some classes of potentials with unbroken SUSY
and shape invariance (SI) which is one of the standard way  and it
is known that the potential algebras of these systems have been
investigated to find exact solutions \cite{Witten, Cooper1, Gend,
infeld, gursey, sukhatme, rasin, balantekin, Lev, Cooper}. SUSY QM
methods and relativistic extensions have been used by many authors
\cite{al, has, chun, gh}.

The purpose of the present paper is to explore new relativistic
complex vector potentials of the non-Hermitian bosonic
Hamiltonians which may be unsolvable and map them into a solvable
but real effective potentials. In the literature,
bosonic/fermionic Hamiltonians with two mode have physical
importance such as Jaynes-Cummings model in solid-state physics
\cite{Jan}, Bose-einstein condensate \cite{sanz}, squeezed states
in a condensate of ultracold bosonic atoms confined by a
double-well potential \cite{sq}.

Using the methods of SUSY QM, we have obtained solutions of
complex vector potentials and showed that in Dirac equation,
decomposing the the vector potential into the real and imaginary
parts leads to derive both exactly and conditionally exactly
solvable potentials. The paper is organized as follows: In section
$2$ a non-Hermitian Hamiltonian model is introduced by us and
mapped into its Hermitian form. Shape invariance which is one of
the effective tool in SUSY QM is given shortly in section $3$.
Section $4$ includes the mapping of the Dirac equation into a
Schr\"{o}dinger-like equation and obtaining new complex and
effective potentials with their exact solutions.
\section{The Non-Hermitian Model and Hermitian Equivalents}
Previous works by the authors have included many aspects of a
non-Hermitian $su(2)$ Hamiltonian known as Swanson Hamiltonian
\cite{swanson, sinharoy, jones, midya, assis, yesiltas}. The
Swanson Hamiltonian is given by
$\hat{H}=\omega(\hat{a}^{\dag}\hat{a}+1/2)+\alpha
\hat{a}^{2}+\beta \hat{a^{\dag}}^{2}$,  where $\hat{a},
\hat{a^{\dag}}$ are annihilation and creation operators, $\omega,
\alpha$ and $ \beta $ are real constants. In this paper, let us
consider a $\mathcal{PT}$ symmetric non-Hermitian model with two
parameters given by
\begin{equation}\label{1}
  \hat{\mathcal{H}}=\omega (\hat{b}^{\dag}\hat{b}+\frac{1}{2})+  \alpha (\hat{b}^{2}-(\hat{b}^{\dag})^{2})
\end{equation}
where  $\dag$ is Hermitian adjoint, $\hat{b}$ is the annihilation
operator given in a general form,
\begin{equation}\label{2}
  \hat{b}=A(x)\frac{d}{dx}+B(x)
\end{equation}
and $A(x)$, $B(x)$ are real functions. $\mathcal{PT}$ operator has
an effect as $x\rightarrow -x$, $p\rightarrow p$ and $i\rightarrow
-i$ in the Hamiltonian, if the operators are taken as
$\hat{b}=\frac{\omega}{2}\hat{x}+\frac{i}{2\omega}\hat{p}$, it can
be seen that the Hamiltonian is $\mathcal{PT}$ symmetric. Now, in
terms of differential operators, (\ref{1}) becomes
\begin{equation}\label{3}
\begin{split}
  \hat{\mathcal{H}}&=-\omega A(x)^{2}\frac{d^{2}}{dx^{2}}+(4\alpha A(x)B(x)-2\omega A(x)A(x)^{'})\frac{d}{dx}\\-
  &(\omega-2 \alpha) A(x)B(x)^{'}-(\omega- 2 \alpha) A(x)^{'} B(x)+\omega B(x)^{2}-\alpha (A(x)A(x)^{''}+(A(x)^{'})^{2})+\frac{\omega}{2}.
  \end{split}
\end{equation}
We may write the eigenvalue equation for (\ref{1}) as given below
\begin{equation}\label{ev}
   \hat{\mathcal{H}} \psi=\varepsilon \psi.
\end{equation}
Here, the pseudo-Hermitian Hamiltonian (\ref{3}) can be mapped
into a Hermitian operator form by using a mapping function $\rho$
\begin{equation}\label{f1}
  h=\rho \hat{\mathcal{H}} \rho^{-1}
\end{equation}
where
\begin{equation}\label{rho}
  \rho=e^{-\frac{2\alpha}{\omega}\int dx \frac{B(x)}{A(x)}}.
\end{equation}
Here we note that $h \psi=\varepsilon \psi$, $\psi=\rho^{-1}\xi$.
So we can introduce operator $h$ which is Hermitian equivalent of
$\mathcal{H}$ as
\begin{equation}\label{f2}
 h= -\omega\frac{d}{dx}A(x)^{2}\frac{d}{dx}+U_{eff}(x)
\end{equation}
here $U_{eff}(x)$ takes the form
\begin{equation}\label{f3}
  U_{eff}(x)=\frac{\omega}{2}-\omega(A(x)B(x))^{'}-\alpha \left((A^{'}(x))^{2}+A(x) A^{''}(x)\right)+(\omega+\frac{4\alpha^{2}}{\omega})B^{2}(x)
  \end{equation}
where the primes denote the derivatives. Then (\ref{f2}) can be
mapped into a Schr\"{o}dinger-like form by using
 \begin{equation}\label{ev2}
   \xi(x)=\frac{1}{A(x)} \Phi(x)
 \end{equation}
Hence, Schr\"{o}dinger-like equation becomes
\begin{equation}\label{f5}
  -\Phi^{''}(x)+\left(\frac{\omega/2-\varepsilon}{\omega A^{2}(x)}-\frac{(A(x)B(x))^{'}}{A^{2}(x)}+
  \frac{\omega^{2}+4\alpha^{2}}{\omega^{2}}\frac{B^{2}(x)}{A^{2}(x)}
  +\frac{\omega-\alpha}{\omega}\frac{A^{''}(x)}{A(x)}-\frac{\alpha}{\omega}\frac{(A^{'}(x))^{2}}{A^{2}(x)}\right) \Phi=0.
\end{equation}
\section{Shape Invariance}
It is very well-known that a quantum system having a
square-integrable ground-state with finite/infinite discrete
energy levels $E_{0}<E_{1}<E_{2}<...$ where the ground-state
energy is chosen to be zero $E_{0}=0$ is a fundamental idea in
supersymmetric quantum mechanics. Generally we can denote the
positive semi-definite Hamiltonian by $\mathbb{H}$ can be given in
a factorized form \cite{Cooper}:
\begin{equation}\label{si1}
  \mathbb{H}=\mathcal{A^{\dag}}\mathcal{A}=-\frac{d^{2}}{dx^{2}}+v(x)
\end{equation}
\begin{equation}\label{si2}
  \mathcal{A}=\frac{d}{dx}-W(x), ~~~~\mathcal{A^{\dag}}=-\frac{d}{dx}+W(x)
\end{equation}
\begin{equation}\label{si3}
  v^{\pm}(x)=W^{2}(x)\pm W^{'}(x).
\end{equation}
We used the unit system $\hbar=2m=1$. Here $W(x)$ is the function
which is real and smooth known as the superpotential and the
ground-state wave-function $\zeta_{0}(x)=e^{-\int^{x} dy W(y)}$ is
nodeles. It is noted that $\mathcal{A}\zeta_{0}(x)=0$. In this
approach, potential depends on a set of parameters $a=(a_{0},
a_{1}, a_{2},...)$ to be expressed by $W(x,a), \mathcal{A}(a),
E(a),...$ The shape invariance condition is
\begin{equation}\label{si4}
  \mathcal{A}(a)\mathcal{A^{\dag}}(a)=\mathcal{A^{\dag}}(a+\Delta)\mathcal{A}(a+\Delta)+E_{1}(a)
\end{equation}
in which $\Delta$ is the shift of the parameters. The entire set
of discrete eigenvalues and corresponding eigenfunctions are
$E_{n}(a)$ and $\zeta_{n}(x,a)$ can be written as
\begin{eqnarray}\label{si5}
  E_{n}(a) &=& \sum^{n-1}_{k=0} E_{1}(a+k \Delta) \label {si6}\\
  \zeta_{n}(x,a) &\sim& \mathcal{A^{\dag}}(a) \mathcal{A^{\dag}}(a+\Delta)...\mathcal{A^{\dag}}(a+(n-1)\Delta) e^{-\int^{x} dy W(y,a+n\Delta)}.
\end{eqnarray}
\section{Dirac Equation}
The Dirac equation which plays an important role in relativistic
quantum mechanics describes relativistic effects due to the speed
and spin of particles. The one dimensional time independent Dirac
equation with effective mass $M(x)$ and vector potential $V(x)$ is
\begin{equation}\label{d1}
  (\hat{\alpha}.\vec{p}+\hat{\beta} M(x)+V\hat{I})\Psi(x)=E\hat{I}\Psi(x)
\end{equation}
where $\Psi$ is the two component spinor wave-function, $E$ is the
energy, $\vec{p}$ is the momentum operator, $M(x)$ denotes the
position dependent mass and $\hat{\alpha}$ and $\hat{\beta}$ are
$2x2$ Dirac matrices in standard representation and $\hbar=c=1$
atomic units are chosen. Let us show the upper and lower
components by $\phi(x)$ and $\theta(x)$. Using
$\alpha=\sigma_{3}$, $\beta=\sigma_{1}$, where $\sigma_{1}$ and
$\sigma_{3}$ are Pauli matrices, and multiplying (\ref{d1}) by
$\sigma_{1}$, then we obtain \cite{Jia1}
\begin{eqnarray*}
  -i \frac{d \theta}{dx}+(E-V(x))\theta-M(x)\phi &=& 0 \\
  i \frac{d\phi}{dx}+(E-V(x))\phi-M(x)\theta &=& 0.
\end{eqnarray*}
If we terminate $\theta$ in above coupled differential equations,
we obtain
\begin{equation}\label{d2}
  -\frac{d^{2}\phi}{dx^{2}}+\frac{1}{M(x)}\frac{dM(x)}{dx}\frac{d\phi}{dx}+\left(2 E V(x)-V(x)^{2}-i \frac{dV(x)}{dx}-i \frac{1}{M(x)}\frac{dM(x)}{dx}(E-V(x))\right)\phi=(E^{2}-M(x)^{2})\phi.
\end{equation}
We use a transformation of the upper component wave-function which
is $\phi(x)=\sqrt{M(x)}\varphi(x)$ in (\ref{d2}), we find that
\begin{equation}\label{d3}
  -\frac{d^{2}\varphi}{dx^{2}}+V_{eff}(x)\varphi=E^{2}\varphi.
\end{equation}
Here, effective potential $V_{eff}(x)$ reads
\begin{equation}\label{ara}
\begin{split}
  V_{eff}(x)&=-V^{2}(x)-i \frac{dV(x)}{dx}+M^{2}(x)+i \frac{V(x)}{M(x)} \frac{dM(x)}{dx}+E \left(2V(x)-\frac{i}{M(x)}\frac{dM(x)}{dx}\right)\\-&\frac{1}{2M(x)}\frac{d^{2}M(x)}{dx^{2}}+\frac{3}{4}\left(\frac{1}{M(x)}\frac{dM(x)}{dx}\right)^{2}.
  \end{split}
\end{equation}
Now we decompose the vector potential $V(x)$ into the real and
imaginary parts in (\ref{ara}) as
\begin{equation}\label{vx}
  V(x)=V_{R}(x)+ i V_{I}(x)
\end{equation}
which leads to
\begin{equation}\label{leads}
\begin{split}
  V_{eff}(x)&=-V^{2}_{R}(x)+V^{2}_{I}(x)+M^{2}(x)+2EV_{R}(x)-\frac{M^{''}(x)}{2M(x)}+\frac{3}{4}\left(\frac{M^{'}(x)}{M(x)}\right)^{2}
  +V^{'}_{I}(x)-\frac{M^{'}(x)}{M(x)}V_{I}(x)\\+&i\left(-2V_{I}(x)V_{R}(x)+2EV_{I}(x)-V^{'}_{R}(x)+\frac{M^{'}(x)}{M(x)}V_{R}(x)-E\frac{M^{'}(x)}{M(x)}\right).
\end{split}
\end{equation}
We may terminate the imaginary part of $V_{eff}(x)$ by using
\begin{equation}\label{im}
  V_{I}=\frac{M(x)^{'}}{2M(x)}+\frac{V_{R}^{'}(x)}{2(E-V_{R}(x))}.
\end{equation}
Because we have obtained a real effective potential expression for
the non-Hermitian Hamiltonian in the last section. Now, we can
give $V_{eff}(x)$ in the form of
\begin{equation}\label{d4}
  V_{eff}(x)=-V_{R}(x)^{2}+M(x)^{2}+2EV_{R}(x)+ \frac{3 (V_{R}(x)^{'})^{2}}{4(E-V_{R}(x))^{2}}+\frac{V_{R}^{''}}{2(E-V_{R}(x_{}))}.
\end{equation}
In order to compare $U_{eff}(x)$ and $V_{eff}(x)$, we may choose
$M(x)$ and $V_{R}(x)$ as
\begin{eqnarray}\label{lab}
  M(x) &=& m_{1}\frac{A^{'}(x)}{A(x)}+m_{2}\frac{B(x)}{A(x)} \\
  V_{R}(x) &=& E-\frac{E}{A(x)}
\end{eqnarray}
and put in (\ref{d4}) where $m_{1}$ and $m_{2}$ are real
constants. Thus, we give another ansatze for $B(x)$ as
\begin{equation}\label{bx}
  B(x)=\gamma A(x)+\beta A^{'}(x)
\end{equation}
where $\gamma$ and $\beta$ are real constants. Afterwards,
$V_{eff}(x)$ takes the form given below:
\begin{equation}\label{Ve}
  V_{eff}(x)=-\frac{E^{2}}{A(x)^{2}}+m^{2}_{2}\gamma^{2}+2\gamma m_{2}\left( m_{1} +\beta m_{2}\right)\frac{A(x)'}{A(x)}+\left((m_{1}+\beta m_{2})^{2}-\frac{1}{4}\right)\left(\frac{A(x)^{'}}{A(x)}\right)^{2}+\frac{A(x)^{''}}{2A(x)}.
\end{equation}
This time, we shall use   (\ref{bx}) in  (\ref{f5}) so that we
would compare (\ref{Ve}) and (\ref{f5}), then we obtain
\begin{equation}\label{d}
\begin{split}
  -\Phi^{''}(x)+[\frac{\omega^{2}+4\alpha^{2}}{\omega^{2}}\gamma^{2}+\varepsilon+\frac{\omega/2-\varepsilon}{A(x)^{2}}+
  \left(\beta^{2}\frac{\omega^{2}+4\alpha^{2}}{\omega^{2}}-\beta-\frac{\alpha}{\omega}\right)\frac{(A(x)^{'})^{2}}{A(x)^{2}} \\+\left(\frac{\omega-\alpha}{\omega}-\beta\right)\frac{A(x)^{''}}{A(x)}+2\gamma\left(\frac{\omega^{2}+4\alpha^{2}}{\omega^{2}}\beta-1\right)
  \frac{A(x)^{'}}{A(x)}] \Phi(x)=\varepsilon \Phi(x)
  \end{split}
\end{equation}
and we can also give $U_{eff}(x)$ as
\begin{equation}\label{Ue}
\begin{split}
    U_{eff}(x)&=\frac{\omega^{2}+4\alpha^{2}}{\omega^{2}}\gamma^{2}+\varepsilon+\frac{\omega/2-\varepsilon}{A(x)^{2}}+
  \left(\beta^{2}\frac{\omega^{2}+4\alpha^{2}}{\omega^{2}}-\beta-\frac{\alpha}{\omega}\right)\frac{(A(x)^{'})^{2}}{A(x)^{2}} \\+& \left(\frac{\omega-\alpha}{\omega}-\beta\right)\frac{A(x)^{''}}{A(x)}+2\gamma\left(\frac{\omega^{2}+4\alpha^{2}}{\omega^{2}}\beta-1\right)
  \frac{A(x)^{'}}{A(x)}.
  \end{split}
\end{equation}
 Hence, we can compare and (\ref{Ue}) and (\ref{Ve}), then we find this set of equations
 \begin{eqnarray}\label{toplu1}
   \varepsilon &=& \gamma^{2}m^{2}_{2}-\frac{\omega^{2}+4\alpha^{2}}{\omega^{2}}\gamma^{2} \\ \label{toplu2}
   \beta &=& \frac{\omega-2\alpha}{2\omega} \\ \label{toplu3}
   -E^{2} &=& \frac{\omega}{2}-\varepsilon \\ \label{toplu4}
   m_{2}(m_{1}+\beta m_{2})&=&\frac{\omega^{2}+4\alpha^{2}}{\omega^{2}}\beta-1. \label{toplu5}
 \end{eqnarray}
 From the last relation we can find
 \begin{equation}\label{m1}
    m_{1} = \frac{1}{2\omega} (-\beta \omega m_{2}\pm\sqrt{\omega^{2}(1+\beta^{2} m^{2}_{2})-4\alpha \omega})
 \end{equation}
 and then, we can give $E$ in terms of parameters $\omega$, $\alpha$ as
 \begin{equation}\label{ekare}
   E^{2}=\frac{\omega}{2}-\gamma^{2}\left(m^{2}_{2}-\frac{\omega^{2}+4\alpha^{2}}{\omega^{2}}\right).
 \end{equation}
 Now we will give two potential models:
\subsection{Example 1: non-$\mathcal{PT}$ symmetric vector potential }
Using some special values of $A(x)$ may give rise to solvable
effective potential models. For instance, if $A(x)=\delta \cosh x$
is chosen, one obtains
\begin{equation}\label{sec}
    V(x)=E-E \sec h x+\frac{i}{2} \frac{\sec h x}{\mu \cosh x+\sin h x}
\end{equation}
that is not a solvable non-$\mathcal{PT}$ symmetric potential, at
the same time, the mass expression is given by
\begin{equation}\label{kutle1}
    M(x)=m_{2}\gamma+m^{-1}_{2}\left(\frac{\omega^{2}+4\alpha^{2}}{\omega^{2}}\beta-1\right)\tan h x.
\end{equation}
In this case $V_{eff}(x)$ is obtained as
\begin{equation}\label{f7}
  V_{eff}(x)=E^{2}-\left(E^{2}-\frac{1}{4}+(m_{1}+\beta m_{2})^{2}\right)\sec hx^{2}+2\gamma m_{2}(m_{1}+\beta m_{2}) \tanh x+
  \gamma^{2}m^{2}_{2}+\frac{1}{4}+(m_{1}+\beta m_{2})^{2}.
\end{equation}
We can give (\ref{f7}) in terms of $\omega$ and $\alpha$ constants
by the aid of (\ref{toplu1})-(\ref{toplu5}):
\begin{equation}\label{f08}
    V_{eff}(x)=V_{0}-V_{1} \sec h^{2} x +V_{2} \tan h x, ~~~~-\infty < x < \infty
\end{equation}
where
\begin{eqnarray}
  V_{0} &=& \frac{\omega}{2}+\gamma^{2}\sigma+\frac{1}{4}+\left(\frac{\sigma \beta-1}{m_{2}}\right)^{2},
   ~~~~\sigma=\frac{\omega^{2}+4\alpha^{2}}{\omega^{2}}\\
  V_{1} &=& \frac{\omega}{2}-\gamma^{2}(m^{2}_{2}-\sigma)-\frac{1}{4}+\frac{(\sigma \beta-1)^{2}}{m^{2}_{2}} \\
  V_{2} &=& 2\gamma \left(\sigma \beta-1\right)
\end{eqnarray}
If we remember the form of the Schr\"{o}dinger-like equation which
is
\begin{equation}\label{f09}
    -\varphi^{''}+V_{eff} \varphi=\bar{E}\varphi, ~~~~\bar{E}=E^{2}-V_{0},
\end{equation}
thus, we would write the ground-state wave-function in terms of
super-potential $W(x)$ as
\begin{equation}\label{vra}
  \varphi_{0}(x)=exp(-\int^{x} W(y) dy).
\end{equation}
We shall put the super-potential in the form of
\begin{equation}\label{W}
    W(x)=C_{1}+C_{2} \tanh x
\end{equation}
where $C_{1}$, $C_{2}$ are constants, using this relation we
obtain the ground-state wave-function $\varphi_{0}(x)$ as
\begin{equation}\label{ph}
    \varphi_{0}(x)=e^{-C_{1}x} (\cos h x)^{-C_{2}}.
\end{equation}
There are boundary conditions as $C_{2} > 0$ and $|C_{1}| < C_{2}$
such that $\varphi_{0}(x)\longrightarrow 0$ when $x
\longrightarrow \pm \infty$. The partner potentials can be given
in the following manner:
\begin{equation}\label{p1}
    V^{+}_{eff}(x)=W^{2}(x)+W^{'}(x)=C^{2}_{1}+C^{2}_{2}-(C^{2}_{2}-C_{2}) \sec h^{2} x+V_{2} \tan h x
\end{equation}
and
\begin{equation}\label{p2}
    V^{-}_{eff}(x)=W^{2}(x)-W^{'}(x)=C^{2}_{1}+C^{2}_{2}-(C^{2}_{2}+C_{2}) \sec h^{2} x+V_{2} \tan h x.
\end{equation}
If we show the ground state energy with $\bar{E_{0}}$, we may give
the expression as below
\begin{equation}\label{ifade}
  W^{2}(x)-W^{'}(x)=-V_{1} \sec h ^{2}x+V_{2} \tan h x-\bar{E_{0}}.
\end{equation}
Now, we can match (\ref{p2}) with (\ref{f08}), one gets
\begin{eqnarray}
  C^{2}_{1}+C^{2}_{2} &=& -\bar{E_{0}} \\
  C_{2}+C^{2}_{2} &=& V_{1} \\
  2C_{1}C_{2} &=& V_{2}.
\end{eqnarray}
Solving these equations, we obtain $C_{1}, C_{2}, \bar{E_{0}}$ as
follows
\begin{equation}\label{c2}
    C_{2}=\frac{1}{2}(-1\pm \sqrt{1+4 V_{1}})
\end{equation}
and we must choose the positive sign in (\ref{c2}) because of the
boundary conditions, this also leads to $V_{1}>0$. The other
constant $C_{1}$ is given by
\begin{equation}\label{c1}
    C_{1}=\frac{2V_{2}}{-1+\sqrt{1+4V_{1}}}.
\end{equation}
and
\begin{equation}\label{ener}
    -\bar{E_{0}}=\frac{1}{4}(-1+\sqrt{1+4V_{1}})^{2}+\frac{V^{2}_{2}}{(-1+\sqrt{1+4V_{1}})^{2}}.
\end{equation}
It is seen that two partner potentials satisfy the well-known
shape invariant relationship
\begin{equation}\label{shape}
    V^{+}_{eff}(x;a_{0})=V^{-}_{eff}(x;a_{1})+R(a_{1})
\end{equation}
where $a_{0}=C_{2}$ and $a_{1}=C_{2}-1$. The reminder $R(a_{1})$
is not depend on $x$ and it contributes to the energy spectrum as
\begin{equation}\label{zero}
    \bar{E_{0}}^{-}=0
\end{equation}

\begin{equation}\label{n}
\begin{split}
    \bar{E}^{-}_{n}&= \sum^{n}_{k=1} R(a_{k})\\
             &= \frac{V^{2}_{2}}{4C^{2}_{2}} +C^{2}_{2}-\frac{V^{2}_{2}}{4(C_{2}-n)^{2}}+(C_{2}-n)^{2}, ~~n=0,1,2,...
\end{split}
\end{equation}
Eventually, using (\ref{ener}) we obtain the relativistic energy
spectrum for (\ref{sec}) as
\begin{equation}\label{sen}
    E_{n}=\pm\sqrt{V_{0}-\frac{V^{2}_{2}}{4\left(-\frac{1}{2}+\frac{1}{2}\sqrt{1+4V_{1}}-n\right)^{2}}+
    \left(\frac{1}{2}\left(-1+\sqrt{1+4V_{1}}\right)-n \right)^{2}}.
\end{equation}
For real energies $1+4V_{1}$ must be positive, i.e.
\begin{equation}\label{sm}
2\omega+4\gamma^{2}(\sigma-m^{2}_{2})+\frac{4(\sigma
\beta-1)^{2}}{m^{2}_{2}}>0
\end{equation}
and
\begin{equation}\label{smm}
    V_{0} >
    \frac{V^{2}_{2}}{4\left(-\frac{1}{2}+\frac{1}{2}\sqrt{1+4V_{1}}-n\right)^{2}}-
    \left(\frac{1}{2}\left(-1+\sqrt{1+4V_{1}}\right)-n \right)^{2}
\end{equation}
Hereafter we shall find the wave-function $\varphi(x)$. In that
case, using (\ref{n}) in (\ref{f09}) we obtain
\begin{equation}\label{denk}
    -\varphi^{''}+\left(V_{2} \tan h x-(C_{2}+C^{2}_{2}) \sec h^{2} x\right)\varphi=\left((C_{2}-n)^{2}-\frac{V^{2}_{2}}{4(C_{2}-n)^{2}}\right)\varphi
\end{equation}
and if we use a new variable $z=-\tan h x$ in above equation and
writing the function as
\begin{equation}\label{fun}
    \varphi=\left(\frac{1-z}{2}\right)^{-r}\left(\frac{1+z}{2}\right)^{-s} P(z)
\end{equation}
then we get
\begin{equation}\label{Pz}
    (1-z^{2})P^{''}(z)+(-2s+2r-(2-2r-2s)z)P^{'}(z)+n(n-2r-2s+1)P(z)=0
\end{equation}
where
\begin{eqnarray}
  r &=& \frac{1}{2}\left(n+\frac{1}{2}(1-\sqrt{1+4V_{1}})-\frac{V_{2}}{2}\frac{1}{n+\frac{1}{2}(1-\sqrt{1+4V_{1}})}\right) \\
  s &=& \frac{1}{2}\left(n+\frac{1}{2}(1-\sqrt{1+4V_{1}})+\frac{V_{2}}{2}\frac{1}{n+\frac{1}{2}(1-\sqrt{1+4V_{1}})}\right).
\end{eqnarray}
Thus the unnormalised wave function and upper spinor component
$\phi_{n}(x)$ are given by
\begin{equation}\label{dfonk}
    \varphi_{n}(x)=\left(\frac{1+\tan h x}{2}\right)^{-r}\left(\frac{1-\tan h x}{2}\right)^{-s} P^{(-2r,-2s)}_{n}\left(-\tan h x\right)
\end{equation}

\begin{equation}\label{wf}
    \phi_{n}=\sqrt{m_{2}\gamma+m^{-1}_{2}\left(\frac{\omega^{2}+4\alpha^{2}}{\omega^{2}}\beta-1\right)\tan h x} \left(\frac{1+\tan h x}{2}\right)^{-r}\left(\frac{1-\tan h x}{2}\right)^{-s} P^{(-2r,-2s)}_{n}\left(-\tan h x\right)
\end{equation}
where $P^{(-2r,-2s)}_{n}\left(-\tan h x\right)$ are the Jacobi
polynomials. In addition to the results here, in \cite{santos,
Jia1} the authors obtained the spectrum of the Dirac equation with
scalar, vector and pseudoscalar potentials. Our results are
consistent with \cite{santos, Jia1} in case of $V_{2}\rightarrow
iV_{2}$.
\subsection{Example 2: non-$\mathcal{PT}$ symmetric vector potential}
The choice of $A(x)=\delta \coth c x$ gives a non-$\mathcal{PT}$
symmetric potential which is given by
\begin{equation}\label{vx}
    V(x)=E-\frac{E}{\delta} \tanh c x+i \left(-\frac{c}{2} \csc h cx \sec h cx+\frac{2c^{2}(m_{1}+m_{2}\beta)\cot h 2cx}{-2c(m_{1}+m_{2}\beta)+m_{2}
    \gamma \sinh 2 c x}\right)
\end{equation}
where $E$ was given in (\ref{ekare}). And the mass expression
reads
\begin{equation}\label{kutle2}
    M(x)=m_{2}\gamma-2m^{-1}_{2} \csc h 2x.
\end{equation}
Thus, $A(x)$, $V(x)$ and $M(x)$ yields the effective potential
given below
\begin{equation}\label{f8}
\begin{split}
  V_{eff}(x)&= E^{2}+c^{2} \sec h^{2} cx \left(1+\frac{E^{2}}{\delta^{2}c^{2}}+(3/4+(m_{1}+m_{2}\beta)^{2})\csc h^{2} cx \right)\\-& 2c\gamma m_{2}
  (m_{1}+m_{2}\beta) \csc h cx \sec h cx-\frac{E^{2}}{\delta^{2}}+\gamma^{2}m^{2}_{2}.
  \end{split}
\end{equation}
Let us take $\beta$ as
\begin{equation}\label{f9}
  \beta=-\frac{m_{1}}{m_{2}}
\end{equation}
to terminate the term $\csc h cx \sec h cx$ in (\ref{f8}), then
(\ref{f8}) turns into
\begin{equation}\label{f10}
  V_{eff}(x)=E^{2}+c^{2} \sec h^{2} cx \left(1+\frac{E^{2}}{\delta^{2} c^{2}}\right)+\frac{3}{4} c^{2} \sec h^{2} cx \csc h^{2} cx-\frac{E^{2}}{\delta^{2}}+\gamma^{2} m^{2}_{2}.
\end{equation}
To obtain a solvable effective potential, we shall add and
subtract $\frac{3}{4} c^{2} \sec h^{2} c x$ to (\ref{f10}), we
obtain
\begin{equation}\label{f11}
  V_{eff}(x)=E^{2}\left(1-\frac{1}{\delta^{2}}\right)+c^{2}\left(\frac{1}{4}+\frac{E^{2}}{\delta^{2}c^{2}}\right)
  \sec h^{2} cx+\frac{3}{4} c^{2} \csc h^{2} cx+\gamma^{2}m^{2}_{2}, ~~~~~0< x < \infty.
\end{equation}
It is reminded that $V(x)$ turns into
\begin{equation}\label{y6}
    V(x)=E-\frac{E}{\delta} \tanh c x-i \frac{c}{2} \csc h cx \sec h cx.
\end{equation}
Next, we shall give the super-potential in this form
\begin{equation}\label{sp}
    W(x)=A \tan h c x-B \cot h c x
\end{equation}
then we obtain the partner potentials and ground state
wave-function as
\begin{equation}\label{bir}
    W^{2}(x)-W^{'}(x)=V^{-}_{eff}(x) = (A-B)^{2}+B(B-c)\csc h^{2} c x-A(A+c) \sec h^{2} c x
\end{equation}
\begin{equation}\label{iki}
    W^{2}(x)+W^{'}(x)=V^{+}_{eff}(x) = (A-B)^{2}+B(B+c)\csc h^{2} c x-A(A-c) \sec h^{2} c x
\end{equation}
and
\begin{equation}\label{ts}
\varphi_{0}(x)=  \left(\cos h c x \right)^{-\frac{A}{c}}
\left(\sin h c x \right)^{\frac{B}{c}}
\end{equation}
here $\frac{A}{c} > 0$ and $\frac{B}{c}>0$ is taken owing to the
boundary conditions. Now, let us compare (\ref{bir}) and
(\ref{f11}),
\begin{eqnarray}
  (A-B)^{2} &=& E^{2}\left(1-\frac{1}{\delta^{2}}\right)+\gamma^{2}m^{2}_{2} \\
  B(B-c) &=& \frac{3}{4}c^{2} \\
  A(A+c) &=& -c^{2}\left(\frac{1}{4}+\frac{E^{2}}{\delta^{2}c^{2}}\right)
\end{eqnarray}
hence we obtain $B=\frac{3c}{2}$,
$A=\frac{c}{2}-\frac{\omega/2-\gamma^{2}(m^{2}_{2}-\sigma)+\delta^{2}m^{2}_{2}}{4c}$.
Shape invariance relation is written as
\begin{equation}\label{s?}
    V^{+}_{eff}(x,a_{0})=V^{-}_{eff}(x,a_{1})+R(a_{1})
\end{equation}
where $a_{0}$ and $a_{1}$ are given as $a_{0}=\{A,B\}$ and
$a_{1}=\{A-c,B+c\}$. If we use the expressions
$\bar{E}=E^{2}-V_{0}$, we find
\begin{equation}\label{to}
    \bar{E_{n}}^{-}=\sum^{n}_{k=1} R(a_{k})=(A-B)^{2}-(A-B-2c n)^{2}.
\end{equation}
Finally the following relativistic energy spectrum of (\ref{vx})
equals
\begin{equation}\label{ren}
    E_{n}=\pm \delta \sqrt{(\gamma m_{2})^{2}+(A-B)^{2}-(A-B-2c n)^{2}}
\end{equation}
where the term inside of the square root must be positive owing to
obtaining real energies. Substituting (\ref{to}) in (\ref{f09}) we
obtain
\begin{equation}\label{to9}
  -\varphi^{''}(x)+\left((A-B)^{2}+B(B-c)\csc h^{2} c x-A(A+c) \sec h^{2} c x\right)\varphi(x)=((A-B)^{2}-(A-B-2c n)^{2})\varphi(x)
\end{equation}
and we use a new variable $y=\cos $h$ 2cx$ and we express the
function $\varphi(x)=(1-y)^{B/c}(1+y)^{-A/c}P(y)$, then the above
equation becomes
\begin{equation}\label{dk}
 (1-y^{2}) P^{''}(y)+\left(-A-B-(B-A+1)y\right) P^{'}(y)+n(n+B-A)P(y)=0,
\end{equation}
thus, wave-function is given by in terms of Jacobi Polynomials
$P^{(B/c-1/2;-A/c-1/2)}_{n}(y)$
\begin{equation}\label{dff}
  \varphi_{n}(x)=(1-y)^{B/c}(1+y)^{-A/c} P^{(B/c-1/2;-A/c-1/2)}_{n}(y).
\end{equation}
Hence the upper component reads
\begin{equation}\label{uc}
  \phi_{n}(x)=\sqrt{m_{2}\gamma-2c(m_{1}+m_{2}\beta)\csc h 2cx}(1-\cos h 2 cx)^{B/c}(1+\cos h 2cx)^{-A/c} P^{(B/c-1/2;-A/c-1/2)}_{n}(\cos h 2cx).
\end{equation}
Results are agree with those obtained earlier \cite{sukhatme}.
\section{Conclusion}

In the present work, we have introduced a Hamiltonian model
$\mathcal{H}$ which is in non-Hermitian form and mapped
$\mathcal{H}$ into a physical Hamiltonian $h$. The time
independent Dirac equation with effective mass in one dimension is
related to $h$ and transformed into the Schr\"{o}dinger-like
equation with the new complex vector potentials $V(x)$ which are
(\ref{sec}) and (\ref{y6}) derived using the algebraic methods. In
Ref.\cite{Jia1}, the authors used real or pure imaginary vector
potentials $V(x)$. It is seen that composing $V(x)$ into its real
and imaginary components leads to more general effective
potentials which are the elements of the Schr\"{o}dinger-like
equation. In example 1 and 2, terminating the imaginary part of
the effective potential we have derived hyperbolic Rosen-Morse
II-type solvable effective potential and hyperbolic generalized
P\"{o}schl-Teller potential II potential. We note that the mass
relations for each case are more general. We have obtained the
solutions of these effective potential models using shape
invariance method. We have seen that the real spectrum of the
Hamiltonian given  for solvable potentials cannot be
obtained by using $\beta=-\alpha$ in Swanson Hamiltonian. Thus,
the metric operator which is positive definite for the so called
Hamiltonian can be searched in the next studies.

We have introduced some graphs for the energy eigenvalues with
respect to $m_{2}$. (\ref{sen}) is used in figure 1 and we note
that different values of the parameters can lead to real or pure
imaginary energy. For the red curve, the energy is real for the
chosen parameters but it can be seen that between $m_{2}=4.2145$
and $m_{2}=5.6142$ we have imaginary energy values as $i
0.0565786$ and $i 0.0310165$ for the blue curve. If we compare
these results, we see that when $n$ takes the larger values,
energy may take imaginary values for some specific values of
$m_{2}$. When it comes to the figure 2, we have real energies  for
the chosen parameters but when $n$ becomes larger again, the
energy is imaginary for some values of $m_{2}$ which is $0\leq
m_{2} \leq 1.404$.

\begin{figure}[!htb]
\centering
\includegraphics[scale=.7]{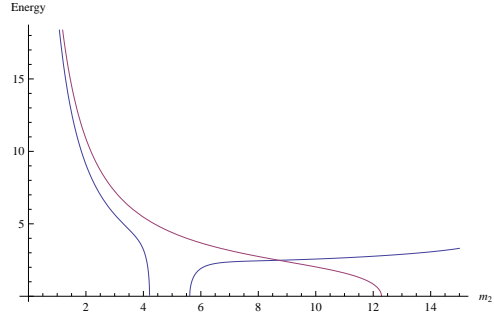}
\caption{Graph of (\ref{sen}) with respect to $m_{2}$, for the red
curve: $n=0, \alpha = 2, \omega = 3, \gamma = 0.1, \beta = 6$; for
the blue curve: $n=3, \alpha = 2, \omega = 3, \gamma = 0.1, \beta
= 6$ } \label{fig:digraph}
\end{figure}

\begin{figure}[!htb]
\centering
\includegraphics[scale=.7]{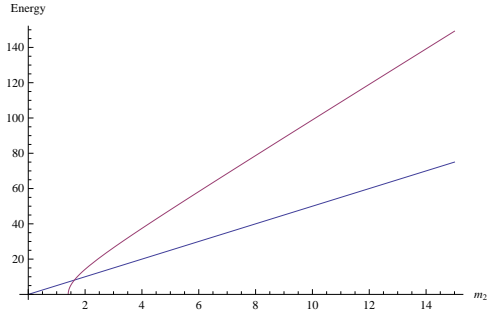}
\caption{Graph of (\ref{ren}) with respect to $m_{2}$, for the red
curve: $n=0, \omega = 5, \alpha = 1, \gamma = 10, \delta= 0.5; c =
3$; for the blue curve: $n=3, \omega = 5, \alpha= 1, \gamma = 10,
\delta = 0.5, c =
 3$} \label{fig:digraph}
\end{figure}

\section*{Acknowledgments}
\medskip
\noindent This paper was written during the authors stay at
Institute of Nuclear Research of the Hungarian Academy of Sciences
(ATOMKI), and the author would also like to thank ATOMKI for its
warm hospitality. Partial financial support of this work under
Grant  from the Higher Education Council of Turkey (Y\"{O}K) is
gratefully acknowledged.


\end{document}